\documentclass[pr1,twocolumn,showpacs,preprintnumbers,reprint,amsmath, amssymb,superscriptaddress]{revtex4-1} 

\usepackage{graphicx} 
\usepackage{dcolumn}  
\usepackage{bm}       
\usepackage{helvet}

\usepackage{mathptmx}

\graphicspath{{FIGS/}}
\begin{document}
\newif\ifnocolor
\nocolortrue

\newcommand{\bea}{\begin{eqnarray}}
\newcommand{\eea}{  \end{eqnarray}}
\newcommand{\bit}{\begin{itemize}}
\newcommand{\eit}{  \end{itemize}}

\newcommand{\be}{\begin{equation}}
\newcommand{\ee}{\end{equation}}
\newcommand{\ra}{\rangle}
\newcommand{\la}{\langle}
\newcommand{\U}{\widetilde{U}}
\newcommand{\brac}[1]{\langle #1|}
\newcommand{\bra}[1]{\langle #1}
\newcommand{\ket}[1]{|#1\rangle}
\newcommand{\ktimes}{\rangle\! \langle}
\newcommand{\op}[2]{|#1\ktimes #2|}
\newcommand{\opoo}{\op{0}{0}}
\newcommand{\opoi}{\op{0}{1}}
\newcommand{\opio}{\op{1}{0}}
\newcommand{\opii}{\op{1}{1}}
\newcommand{\rhoq}{\rho_{\rm q}}
\newcommand{\rhoe}{\rho_{\rm env}}
\newcommand{\env}{{\rm env}}
\newcommand{\qubit}{{\rm qubit}}


\def\bra#1{{\langle#1|}}
\def\ket#1{{|#1\rangle}}
\def\bracket#1#2{{\langle#1|#2\rangle}}
\def\inner#1#2{{\langle#1|#2\rangle}}
\def\expect#1{{\langle#1\rangle}}
\def\e{{\rm e}}
\def\proj{{\hat{\cal P} }}
\def\tr{{\rm Tr}}
\def\H{{\hat H}}
\def\Hdag{{\hat H}^\dagger}
\def\Lop{{\cal L}}
\def\Ehat{{\hat E}}
\def\Edag{{\hat E}^\dagger}
\def\Shat{\hat{S}}
\def\Sdag{{\hat S}^\dagger}
\def\Ahat{{\hat A}}
\def\Adag{{\hat A}^\dagger}
\def\U{{\hat U}}
\def\Udag{{\hat U}^\dagger}
\def\Zhat{{\hat Z}}
\def\Phat{{\hat P}}
\def\Op{{\hat O}}
\def\id{{\hat I}}
\def\x{{\hat x}}
\def\P{{\hat P}}	
\def\Px{\proj_x}
\def\Pr{\proj_{R}}
\def\Pl{\proj_{L}}
\def\ODR{O_{_{\rm DR}}(t)}
\def\ODRn{O_{_{\rm DR}}(n)}
\newcommand{\equa}[1]{Eq.~(\ref{#1})}
\newcommand{\comp}[1]{{\bf #1}}
\newcommand{\mcH}{\mathcal{H}}
\renewcommand{\e}{\text{env}}
\newcommand{\s}{{\rm sys}}
\newcommand{\eref}[1]{Eq.~(\ref{#1})}
\newcommand{\Eref}[1]{Eq.~(\ref{#1})}
\title{ Non-Markovian Quantum Dynamics and Classical Chaos}
\author{Ignacio Garc\'{\i}a-Mata}
\affiliation{Instituto de Investigaciones F\'isicas de Mar del Plata (IFIMAR, CONICET), 
Universidad Nacional de Mar del Plata, Mar del Plata, Argentina.}
\email{i.garcia-mata@conicet.gov.ar}
\affiliation{Consejo Nacional de Investigaciones Cient\'ificas y Tecnol\'ogicas (CONICET), Argentina}                         
\author{Carlos Pineda}
\affiliation{ \mbox{Instituto de F\'isica, Universidad Nacional Aut\'onoma de M\'exico, 
M\'exico D.F., 01000, M\'exico}}
\author{Diego Wisniacki}
\affiliation{\mbox{Departamento de F\'{\i}sica ``J. J. Giambiagi", 
             FCEN, Universidad de Buenos Aires, 1428 Buenos Aires, Argentina}}
\date{\today}

\begin{abstract}
We study the influence of a chaotic environment in the evolution of an open quantum system.
We show that there is an inverse relation between chaos and non-Markovianity.
In particular, we remark on the deep relation of the short time non-Markovian behavior with the revivals of the
average fidelity amplitude -- a fundamental quantity used to measure sensitivity to 
perturbations, and to identify quantum chaos. The long time behavior is established as a 
finite size effect which vanishes for large enough environments.
\end{abstract}

\maketitle
\section{Introduction}
The advent of quantum information and quantum technology has brought us  to
a deeper understanding of the basics of quantum mechanics.  
As a result, various fundamental aspects, such as 
equilibration~\cite{PhysRevLett.108.080402},
simulability~\cite{1751-8121-40-25-S13}, and even
foundations~\cite{PhysRevA.84.012311}, have been revised.  
This framework has provided both 
valuable insight into the 
foundations and new technological achievements.  But theory and its related
experiments  have advanced asymmetrically mainly due to the impossibility to
isolate completely the experimental setup, leaving the system `open' and
exposed to decoherence~\cite{ZurekRMP}. 

Many quantum open systems problems can be solved after
assumptions are made.
The  widely used   Born-Markov approximation, 
successfully applied to describe many physical
situations~\cite{BreuerBook},
is associated with both a
memory-less environment and a weak coupling between system and 
environment. 
However, recently, interest in quantum open systems where this assumption no
longer applies -- usually called non-Markovian (NM) evolution -- has
flourished~\cite{Daffer2004,Breuer2009,Rivas2010}. 

 A natural question to ask is to what extent the dynamical properties 
of the environment can extend
the validity of the Born-Markov approximation,
even beyond the weak coupling regime.
In other words, 
how well does a chaotic environment reproduce Markovian evolution.
In this paper, we address this question analytically and numerically by means
of a probe qubit  coupled to a generic environment where different degrees of
chaos can be tested.  
Exploring the time evolution of non-Markovianity
measures, we show that the stronger the chaos in the
environment, the more Markovian the evolution is.
We build upon previous
knowledge~\cite{garciamaNJP,Haikka2012} of measurable quantities such as the
fidelity amplitude which can be directly related to recently proposed measures
of NM behavior.  
Moreover, we establish  that the
short time behavior of the fidelity amplitude determines the characteristics of
the qualitative NM behavior.  The lingering, long time
contributions, are finite size effects which contribute only as a linear term
with a slope that goes to zero as the environment size goes to infinity.  The
remaining non-Markovianity is thus size independent and due to short time
revivals in the fidelity amplitude decay.

The paper is organized as follows. In Sect.~\ref{sect2} we describe the system we use for our analysis. It consists of a qubit in the presence of an environment whose evolution is subject to the state of the qubit. In 
Sect.~\ref{sect3} we briefly describe the Non-Markovianity measure based on distinguishability and its relation to the fidelity decay of the environment. The numerical results and analysis are 
done in Sect.~\ref{sect4} and we present some concluding remarks in Sect.~\ref{sect5}.
\section{System}
\label{sect2}
We
consider a system-environment situation, whose Hilbert space is
$\mcH=\mcH_\s \otimes \mcH_\e$, where $_\s$  and $_\e$ denote
system and environment. Any Hamiltonian can then be split as 
\begin{equation}
H=H_\s + H_\e + \epsilon V_{\s, \e},
\label{eq:hamiltonian}
\end{equation} 
where $\epsilon$ is real and controls the strength  of the interaction
between  system and  environment. We further
restrict to the case in which $\mcH_\s$ represents a qubit, i.e. when $\dim
\mcH_\s=2$, and we call 
$N=\dim \mcH_\e$.  
The 
choice of  
the terms in \eref{eq:hamiltonian} results in  different physical situations. 
However, instead of specifying the particular form of 
\eref{eq:hamiltonian}, we shall impose two general conditions. 
The first one involves the nature of the
interaction. We shall assume that it is factorizable, i.e. that
$V_{\s, \e} = V_\s \otimes V_\e$. Such structure appears in a wide variety of
situations, including Ising interaction and atom-field interaction under various
approximations \cite{Scully97}. The second assumption 
is to consider that the evolution of the central system, $\mcH_\s$, either
occurs at much smaller time scales  than that 
in which decoherence
occurs or that it is a multiple of $V_\s$. 
In the former, one can safely ignore the contribution to the dynamics,
and in the latter case, $H_\s$ can be included in the interaction term and can  keep
the factorizable structure unaffected.  This occurs, e.g.
in the case of 
a strong magnetic field applied to a set of interacting spins and thus is of
particular importance in, among others, nuclear magnetic resonance (NMR). In this
situation, one can write 
\begin{equation}
H=\opoo\otimes H_0+\opii\otimes H_1
\label{eq:H}
\end{equation}
with $H_0$ and $H_1$ acting only on the environment and both $\opoo$, $\opii$
being projectors onto some orthonormal basis of the qubit. This Hamiltonian has
already been introduced in \cite{Zurek2002} and can be interpreted as
having an environment whose evolution is conditioned by the state of the qubit.
The initial state of the system shall be 
$\rho_{\s, \e}(0)=\rho_\s(0) \otimes \rho_\e$. Notice that
the only condition imposed on the initial state is that it be a
product state.
The evolution of the qubit is 
\begin{equation}
\label{eq:rhot}
\rho_\s(t)=\tr_\e \left[U(t)\rho_\s (0)\otimes \rho_\e U^\dagger(t)\right]
\end{equation}
with 
\begin{equation}
U(t)=\opoo U_0(t)+\opii U_1(t)
\end{equation}
and 
$U_j(t)=\exp(-i t H_j/\hbar)$. 
It is convenient to rephrase \eref{eq:rhot} 
in terms of a quantum channel 
\begin{equation}
\rho_\s(t)=\Lambda(t)[\rho_\s(0)]
\end{equation}
Notice that although 
$\Lambda(t)$ results from tracing out the
environment it still depends on $\rhoe$.

The matrix elements of the channel 
induced by \eref{eq:rhot}, in the Pauli basis are
\begin{equation}
\Lambda^{(t)}_{j,k} = 
	\frac{1}{2} \tr\left[\sigma_j U(t) \sigma_k \otimes \rho_\env 
	U^\dagger(t)\right].
\label{eq:Lambda}
\end{equation}
If we take $\sigma_0 = \mathbb{I}$ and  
$\sigma_{1, 2, 3}=\sigma_{x, y, z}$, 
the channel takes the  
simple form 
\begin{equation}
\Lambda =\begin{pmatrix}
1 & 0 & 0 & 0 \\
0& \Re[f(t)]& \Im[f(t)]&0\\
0& \Im[f(t)]& \Re[f(t)] &0 \\
0 & 0 & 0 & 1
\end{pmatrix}.
\label{eq:paudepha}
\end{equation}
with $f(t)=\tr [\rho_\env U_1(t)^\dagger U_0(t)]$ being the expectation value of the echo
operator. 
  If $\rho_\env$ is a
pure state \cite{Haikka2012} then $|f(t)|^2$ is the Loschmidt echo (LE)
\cite{Jalabert2001} -- also called fidelity -- originally proposed to measure
sensitivity to perturbations in the Hamiltonian as a signature of quantum chaos
\cite{Peres1984}. The LE decays as a function of time and the -- more or less
-- universal decay regimes have been extensively studied (see reviews
\cite{Gorin2006,Jacquod2009}). 
The
environment could in fact be in a pure state -- e.g. in a thermal ground state
at zero temperature. However it is probably easier to imagine
the environment being in a mixed state --
e.g. at thermal equilibrium at a given temperature.  We choose then the
environment to be in the maximally mixed state, i.e. proportional to the
identity.
In that case
we obtain the real and
imaginary part of the average fidelity amplitude (AFA), 
\begin{equation}
\la f(t) \ra = \frac{1}{N} \tr[ U_1(t)^\dagger U_0(t)]
\end{equation}
i.e.  the average value of the echo
operator  with respect to an orthonormal basis. 
We remark that the choice of basis 
(or any complete set)
is arbitrary.
This fact contrasts the case of the LE where
the {\it kind} of states in the set is crucial~\cite{nacho2011}.


\section{Fidelity amplitude measures Non-Markovian behavior} 
\label{sect3}
During a classical Markovian process the distance between two
initial distributions decreases monotonically. Deviations
from this behavior are a landmark of non-Markovianity. 
Breuer
~{\it et al.} \cite{Breuer2009} used this property to {\it define} a measure
of  NM behavior
in a quantum setting. The distance can be chosen as to link non-Markovianity
with distinguishability of states and thus information flow
between the system and its surroundings. Such a
measure is defined as
\begin{equation}
{\cal M}=\max_{\rho_{1,2}(0)}\int_{\sigma>0} dt\sigma(t,\rho_{1.2}(0)),
\end{equation} 
where $\sigma(t,\rho_{1.2}(0))=dD(\rho_1(t),\rho_2(t))/dt $ is the rate of change of the trace distance 
\begin{equation}
D(\rho_1(t),\rho_2(t))=\frac12{\rm tr}|\rho_1(t)-\rho_2(t)|
\end{equation}
between initial states $\rho_{1,2}(0)$. 
In \cite{Rivas2010} two other measures were proposed, based on deviation of
semi-group properties of quantum flows. 
Both study the physicality of the induced instant map at intermediate times, 
one via the Jamio\l kowski isomorphism and the other via the entanglement (as measured
with the concurrence)  with an ancilla qubit.
It is straightforward to show that for channels like \eref{eq:paudepha}
the measure induced by the entanglement is proportional to ${\cal M}$ .

In our case, the states that maximize ${\cal M}$ are 
$\rho_{\pm}=(I\pm(a \sigma_x+b\sigma_y))/2$, with $|a|^2+|b|^2=1$, leading to 
\begin{equation}
	\label{eq:Mconf}
{\cal M}=2 \int_{\dot{|f|}>0} \frac{d|\langle f(t)\rangle|}{dt}.
\end{equation}
In other words, \equa{eq:Mconf} means that NM behavior is directly 
related to the positive derivative of the AFA 
as a function of time \footnote{while in the process of writing this manuscript, 
this relation was independently established in \cite{Haikka2012}}. 

For  fully chaotic systems both the AFA and the LE
saturate around a value that depends on $\hbar$.
After saturation sets in, the state is approximately random and 
the value of both fidelity and fidelity amplitude  fluctuate.
As a consequence, we expect the NM measure to grow indefinitely.  
Thus in our calculations of ${\cal M}$ we modify the original definition in
\cite{Breuer2009} and calculate the NM measure up to a certain time. Regardless,
the measure at time $t$ still holds its meaning, i.e. the larger ${\cal M}(t)$
means the distance between the two states has ceased to decrease (or  increased) 
more in that period of time, which implies a more NM behavior.

\section{Non-Markovianity and chaos:  results}
\label{sect4}
Now we consider the long standing question of the relation between chaos
and Markovianity. To do so
we model the environment using simple but fully
featured  systems: quantum maps on the torus. 
The quantization of the torus implies that both position and 
momentum are discretized and the effective Planck constant is
the inverse of the Hilbert space dimension $N$. 
In this setting, a quantum map is simply a unitary $U$ acting on an
$N$ dimensional Hilbert space.

\begin{figure} 
\includegraphics[width=\linewidth]{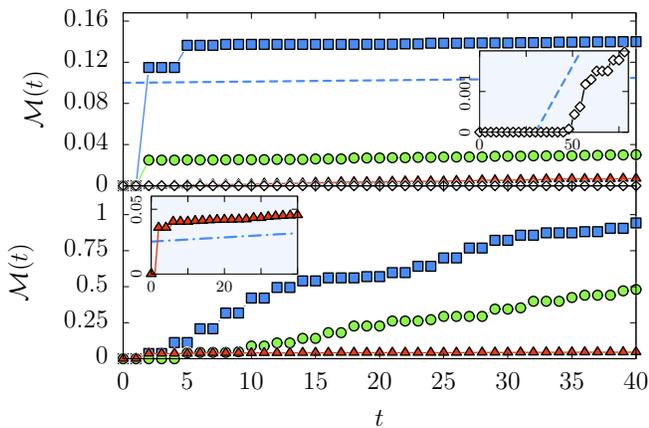}
\caption{(color online)  ${\cal M}(t)$  for the PCM (top) and  the HM (bottom). 
Top: (squares) $\lambda=0.96$, (circles)$ \lambda=1.76$, 
(triangles) $\lambda=5.99$. $\delta K/\hbar=3.635$.
(diamonds) $\lambda=0.96$, $\delta K/\hbar=0.64$ (weak coupling).
Bottom: (squares) $ k=0.001$ (regular), (circles) $ k=0.25$ (mixed),
(triangle) $ k=1$ (fully chaotic).
Top inset: (diamonds) weak coupling case. 
Botom inset: (triangles) $k=1$, note the 
linear dependence.
The slopes of the straight lines are $\alpha=1/2 N$ (top panel, dashed) 
and $\alpha=2/3 N$ (bottom inset, dot-dash).
$N=4096$. \label{fig:uno}}
\end{figure}

We consider two different maps.
First, the quantum perturbed cat map (PCM)
\begin{equation}
\label{eq:pcat}
U_{{\rm c},K}=e^{-i \pi N a \hat{p}^2}e^{i \pi N a \hat{q}^2 } e^{i \pi N K 
(2 \sin (2 \pi \hat{q})- \sin (4\pi \hat{q}))}
\end{equation} 
where $\hat q$ and $\hat p$ are the generators of periodic position and momentum
translations on the torus with discrete eigenvalues $0$, $1/N$, \dots,
$(N-1)/N$.
The subindex $K$ denotes the depth of the kicking potential. 
For $a=1$ and $K=0$ it is the quantum version of 
Arnold's cat map, 
a uniformly hyperbolic and mixing map of the torus onto itself, 
which is a paradigmatic example 
of chaos in two dimensions.
The positive Lyapunov exponent $\lambda$, which determines the rate
of exponential divergence of classical trajectories, is uniform over the whole 
phase space. We explore different 
degrees of chaos by changing $a$
since, for small $K$,  $\lambda\approx \ln ((2+a^2 +\sqrt{a^2(4+a^2)})/2)$.
Here $U(t)\equiv U^t$ where now $t$ is an integer, and $U_0=U_{{\rm c},K}$ and 
$U_1=U_{{\rm c},K+\delta K}$. 

The other map we consider is 
\begin{equation}
\label{eq:harp}
U_{{\rm H}, k,k'}=e^{i N k \cos (2 \pi \hat{q})} e^{i N k' \cos (2 \pi \hat{p})}
\end{equation}
which corresponds to the Harper map (HM) 
 \cite{Leboeuf1990}. It is an approximation of the motion of  
an electron in a crystal under the action of an external field \cite{ArtusoScholar}. 
For $k\lesssim 0.11$, the dynamics described by 
the associated classical map is regular, while for $k\gtrsim 0.63$ 
there are no remaining
visible regular islands. 
We set $U_0 =U_{H,k,k}$ and $U_1 =U_{H,k+\delta k,k}$.
\begin{figure}
\includegraphics[width=\linewidth]{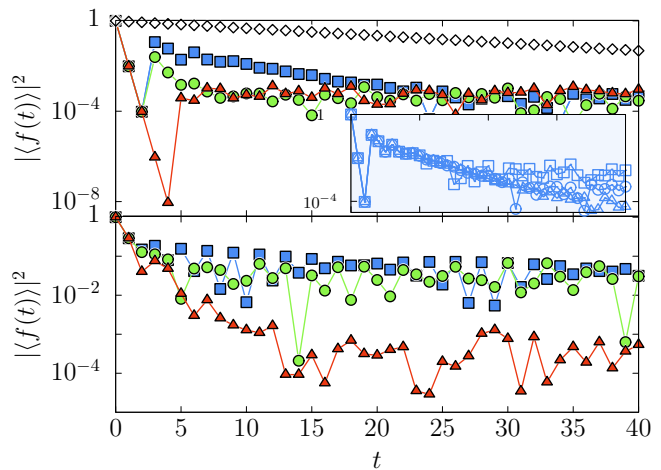}
\caption{(Color online)
$|\langle f(t)\rangle |$, same cases as Fig.~1 for the PCM (top) and  the HM (bottom). 
Inset: $\lambda=0.96$ for the PCM, for  $N=512,4096,\, 16384$.
\label{fig:dos}}
\end{figure}

We now take the result of \equa{eq:Mconf} and compute numerically ${\cal M}(t)$ 
for the two maps.  
Notice that the structure of the maps is $U_1=U_0P(\epsilon)$ (with $\epsilon=\delta K$ or $\delta k$ ).
The $\epsilon\to 0$ limit implies $P(\epsilon) [\mbox{and} f(t)]\to 1$ (i.e. no decoherence).   
The coupling strength
is given by $\delta K$ and $\delta k$. 
For weak couplings, (in the chaotic case) the AFA decays exponentially, 
and the rate depends quadratically on the coupling parameter -- the Fermi golden rule regime (FGR).
Here the evolution is  expected to be Markovian.
Throughout this paper we consider mainly coupling strengths beyond the FGR.

\begin{figure}
\includegraphics[width=0.95\linewidth]{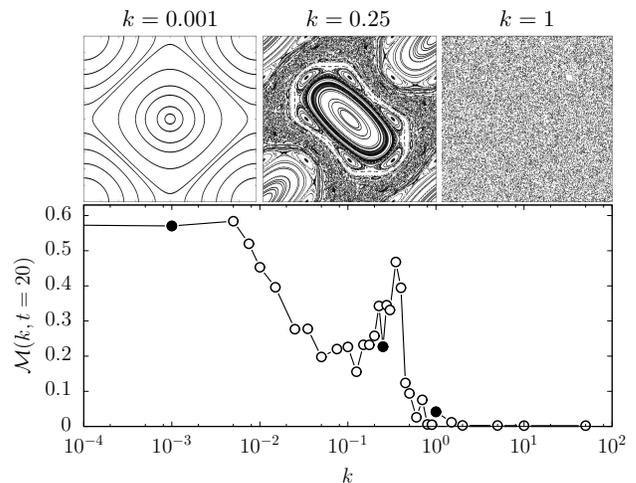} 
\caption{${\cal M}(t=20)$  as a 
function of $k$, for $t=20$, for the HM with  
$N=4096$ and $\delta k=0.00113$.
Top: phase space diagram for
three examples of $k$: regular, $k=0.001$; mixed $k=0.25$; chaotic $k=1$.
The corresponding points in the bottom panel are drawn in solid black. 
\label{fig:mdek}
}
\end{figure}

The AFA is evaluated 
directly by averaging the echo operator 
over a complete set of states.  
For the PCM we change
$a$ in \equa{eq:pcat} so we can assess the change in ${\cal M}$ for different
levels of chaoticity. For the HM varying  $k$ in \equa{eq:harp}, we go from
integrable to completely chaotic.  In Fig.~\ref{fig:uno} (top), we show the NM
measure ${\cal M}(t)$  for three different examples of the PCM 
with varying degrees of chaoticity.
On the bottom of  Fig.~\ref{fig:uno}, we show the same for
the HM, where we 
show the AFA for three different $k$ (regular, mixed, and chaotic).
For the PCM  after a small number of steps, there
appear three distinct jumps.  As expected, the larger $\lambda$ is, the smaller is the
jump, which confirms the intuitive relation that the more chaotic the
environment is,  the more Markovian is the evolution.  After this short time behavior,
the three cases exhibit linear growth  of ${\cal M}(t)$.
The explanation is simple.  For fully chaotic systems,
at a time of the order of Ehrenfest time [$t_{\rm E}=\ln(1/\hbar)/\lambda$], the
AFA saturates -- but oscillates -- around $1/N$  
This saturation corresponds
to the overlap between two completely random states, and is approximately constant.  
This implies  
that ${\cal M}$ will grow linearly and that the slope $\alpha$ will be
proportional to $1/N$.  The proportionality constant depends on the map (for
the PCM, regardless of the value of $\lambda$, we found the slope to be
$\alpha\approx 1/2 N$).  
For completeness, we include the case $\lambda=0.96$ in the weak 
coupling regime. As expected, ${\cal M}(t)=0$  visibly up to $t\approx 50$. After that we have linear growth.
On the bottom of Fig.~\ref{fig:uno}, we show   ${\cal
M}(t)$  for the HM for three qualitatively different cases. In the case in which
the classical dynamics is regular ($k=0.001$), we observe that   up
to the times shown ${\cal M}(t)$ increases non-linearly. 
There is a saturation
of the AFA, but \emph{not} at $1/N$, which eventually leads to a linear growth of ${\cal M}$.
This saturation for small $k$  takes place at much larger times.  When the
dynamics is almost fully chaotic ($k=1$), 
there is a very small jump after which 
there remains only the linear growth due to fluctuations around
the saturation value. The slope of this linear growth
is $\alpha\approx 2/3 N$. 
In the parameter region where the KAM tori of the
HM begin to break, there is a combination between regular and chaotic dynamics (initial states can 
have components inside regular islands and components inside the chaotic sea) and the behavior 
is less intuitive. In fact what is 
observed --in Fig.~\ref{fig:uno} for $k=0.25$ 
--is that the 
environment modeled by a HM in the
transition from regular to chaotic can be strongly non-Markovian [see also
Fig.~\ref{fig:mdek}].

In both situations, the long time behavior 
for the NM measure is linear. This would imply ${\cal M}\to \infty$. 
However, this assertion presents no problems in our analysis. 
The slope of the long time linear regime goes to zero as $N$ grows. 
Intuition suggests  `large' environment as a necessary condition of Markovianity.
However, in the $N\to\infty$ limit, there will always remain the short time 
value attained by  ${\cal M}$ (see Fig.~\ref{fig:uno}), which is independent of
$N$ (Fig. \ref{fig:dos}). 

To shed more light on the 
results displayed in Fig.~\ref{fig:uno} 
we focus 
on the evolution of the AFA as a function of time for the
cases considered above.  In Fig.~\ref{fig:dos} we show examples of the decay of
the square of the AFA. In the top panel, we show results for the PCM for
three different values of $\lambda$ (i.e. different $a$ in \equa{eq:pcat}).  
In the bottom panel, we show the same for the HM, with
three values of $k$ (regular, mixed, and chaotic). 
In contrast with the LE, %
the AFA  is independent of the
type if initial states and decays exponentially with two distinct  decay rates.
The short time decay rate $\Gamma$ can be related to uncorrelated -- random
-- dynamics \cite{garciamaNJP}.
The value of $\Gamma$ can be computed using
semiclassical methods.
This decay rate can diverge, meaning that the short time 
decay can be extremely fast.
These divergences -- which depend on the type of perturbation,
and are more evident the larger $\lambda$ is -- 
could be related to the phenomenon known as 
survival collapse \cite{Horacio} after which the largest revivals appear.
For the numerical results, on the top  we chose a value of $\delta k$ for the PCM
which corresponds to a large $\Gamma$ (near the diverging values), where
the largest revivals have been observed \cite{nacho2011,garciamaNJP}.
We remark, moreover, that the short time decay of the AFA is independent of $N$ and  
therefore so is the revival. In the inset of Fig.~\ref{fig:dos}
we see that the AFA (for the PCM, with $\lambda=0.96$) 
is almost %
equal for three different values of $N$ 
up to $t\approx 10$.	
This is important 
because the short time revivals will provide 
the main contribution to ${\cal M}$. While this contribution 
remains constant with $N$ 
the long time contribution goes to zero as $1/N$.
The curve with diamonds supports the results shown in 
Fig.~\ref{fig:uno} [top] for the weak coupling regime.

The possibility to assess the behavior as an
environment model, by changing one parameter, from regular to chaotic, is
indeed tempting.  In Fig.~\ref{fig:mdek} we computed ${\cal M}$ at a fixed time
for the HM for different values of $k$. We chose $t=20$, around the time in which
the fastest decaying case starts to saturate (see Fig~\ref{fig:dos}, bottom).
We see that for
small $k$ (regular dynamics),  ${\cal M}$ takes a constant value (which, apart
from the fixed time,  depends on $N$ and $\delta K$) and there is a
transition where  ${\cal M}$ depends on $k$ just where the KAM  tori begin to
break. When the dynamics is fully chaotic, the value of  ${\cal M}$ (at $t=20$)
again takes a constant value. Fig.~\ref{fig:mdek} is a clear example of the
expected behavior: regular environments are expected to be more NM while the NM
behavior that appears to linger in the chaotic regime is due to the same
oscillatory behavior around the saturation value mentioned for the case of the
PCM. In the transition region, $0.11\lesssim k \lesssim 0.63$, there is coexistence between
tori and chaotic regions. In the first place, the existence of regular islands
implies that even though there will be leaking -- by tunneling -- to the
chaotic regions, the saturation will take much longer. In addition, the area
occupied by the chaotic region is smaller than the torus, therefore, the
saturation value is larger than $1/N$.  
We have checked for other times (up to $t\sim 1000$) and also other methods (not shown) 
-- e.g. taking as NM value the $y$-intercept of an asymptotic  
linear fit -- and the qualitative behavior is the same. Further studies are needed 
in order to fully grasp the behavior
in the intermediate region.
 
\section{Conclusions}
\label{sect5}
 We addressed the issue of how well a chaotic environment can 
model Markovian evolution.
 We used a probe qubit as a system
coupled to an environment modeled by a quantum map.  In this setting there is a
straightforward relation between some measures of NM behavior and
the AFA. 
The study of the time evolution of the NM measure has shown %
 that the stronger the chaos of the
environment (in the PCM larger $\lambda$), the more Markovian the evolution will
be, even if the coupling is strong.
Furthermore, there are two well defined regimes. For short times, there is no dependence 
with $N$ and the NM is measured by revivals in the AFA. 
In contrast, for large times, the measure grows linearly with a slope that vanishes
as $\propto 1/N$.
Thus, in accordance with \cite{Znidaric2011}, as $N\to\infty$ there can be a remaining non 
vanishing value for non-Markovianity, 
for a chaotic environment. %
The revivals of the LE were recently 
related to NM behavior \cite{Haikka2012}. 
Here we make a more general approach by allowing the bath to be in a thermal state and
expressing non-Markovianity in terms of the AFA -- a quantity which is independent
of the set of states over which the average is done. 

\acknowledgments 
Discussions with P. Haikka and J. Piilo are
acknowledged. C.P. received support from the  projects
CONACyT 57334 and  UNAM-PAPIIT IN117310. I.G.M. and D.A.W.
received support from ANCyPT (PICT 2010-1556), UBACyT,
and CONICET (PIP 114-20110100048 and PIP 11220080100728). 
%

\end{document}